# NMR Tracing of Hyperfine-Mediated Nuclear Spin Diffusion in Fractional Quantum Hall Domain Phases

S. Miyamoto,[1, 2, *] T. Hatano,[1, †] S. Watanabe,[3] and Y. Hirayama[1]

[1]*Department of Physics, Tohoku University, 6-3 Aramaki Aza Aoba, Aoba-ku, Sendai 980-8578, Japan*
[2]*School of Fundamental Science and Technology, Keio University,*
*3-14-1 Hiyoshi, Kohoku-ku, Yokohama 223-8522, Japan*
[3]*Institute of Science and Engineering, Kanazawa University, Kanazawa 920-1192, Japan*
(Dated: May 23, 2016)

We present an enhanced diffusion of nuclear spin polarization in fractional quantum Hall domain phases at $\nu = 2/3$. Resistively-detected NMR mediated by electrically driven domain-wall motion is used as a probe of local nuclear polarization, manifesting pumping-dependent signal saturation behavior. This reveals that a relatively homogeneous polarization profile spreads even to places distant from pinning centers of the domain walls. We attribute this to the fact that the pumped nuclear polarization near the domain walls rapidly diffuses into the domains where nuclei experience Knight fields on comparable levels. The anomalous enhancement of nuclear diffusion may be interpreted in terms of indirect hyperfine-mediated interaction between nuclear spins in the domains.

Following the early work using shallow donors in GaAs bulk [1], controversial findings have shown that hyperfine interaction strongly modifies nuclear spin diffusion in quantum dots [2–6]. A confined electron wavefunction generates nonuniform hyperfine fields that suppress the nuclear spin diffusion via *direct* dipole-dipole coupling [2, 3]. Meanwhile, a virtual process of nuclear spin exchange via the hyperfine interaction couples distant nuclei without flipping the mediating electron spin [4]. Such *indirect* nuclear spin coupling may assist the distribution process of nuclear polarization in the quantum dots [5, 6].

In quantum Hall (QH) systems as well, the hyperfine interaction is evident in the nuclear spin behavior. The strong coupling between nuclear and electron spins has enabled the probing of a wide spectrum of non-trivial electron spin states in the QH systems [7–12]. In particular, the spin domain structures at the fractional Landau level (LL) filling factor $\nu = 2/3$ are often used for efficient dynamic nuclear polarization (DNP) [13–15]. As illustrated in Fig. 1(a), when an energetically balanced state is achieved between the Zeeman energy, $E_Z$, and the Coulomb energy, $E_C$, for composite fermions (CFs), domain walls (DWs) form at the spin phase boundary of unpolarized domains (UDs) and polarized domains (PDs). It is generally accepted that when an electrical current flows along such compressible DW regions, nearby nuclear spins are pumped selectively. The nuclear polarization accumulated near the DWs is then predicted to spread in both in-plane and out-of-plane directions. So far, the current-pumped nuclear polarization in one quantum well (QW) has been detected by another closely-separated QW, which allows the study of the nuclear spin diffusion in the out-of-plane direction [16, 17]. These experiments, however, incorporate little hyperfine modification to nuclear diffusion coefficients since these coefficients are mainly determined in the absence of a two-dimensional electron gas (2DEG).

In this work, we investigate the in-plane profile of nuclear polarization in the domain structures at $\nu = 2/3$, where the hyperfine impact on the nuclear diffusion is significant. Modifying the filling factor via gate electric fields induces spatial oscillation of DWs, by which nuclear magnetic resonance (NMR) is caused by electrical means [18, 19]. This method of nuclear electric resonance (NER) is here utilized for tracing nuclear polarization extended into the domains. As a result, it is demonstrated that around the DWs relatively uniform polarization is established due to rapid nuclear diffusion. This diffusion is stronger than that predicted by *direct* dipole-dipole coupling, but we show that the enhanced diffusion in the domains may be associated with the *indirect* hyperfine-mediated coupling between nuclear spins.

The sample used is a Hall bar with a 50-$\mu$m width which consists of a GaAs/AlGaAs QW structure of width $w = 20$ nm. It is located inside a mixing chamber of a dilution refrigerator to keep the temperature at 100 mK. An electron density $n_e$ is varied by a back gate embedded below the QW. For $n_e = 1.2 \times 10^{15}$ m$^{-2}$, the electron mobility is 175 m$^2$/V·s. In order to achieve the spin phase transition (SPT) around $\nu = 2/3$, a magnetic field of $B = 7.4$ T is applied perpendicular to the 2DEG plane. Each measurement sequence is started by initializing nuclear polarization via skyrmions, at $\nu_{\text{init}} = 0.9$ for the duration of $\tau_{\text{init}}$ [7]. After the filling factor is set to $\nu_{\text{pump}} \sim 2/3$, ac current application at $I_{\text{pump}}$ and 13.4 Hz gives rise to the DNP for the duration of $\tau_{\text{pump}}$. Except as otherwise noted, $\tau_{\text{init}}$ is taken to be equal to $\tau_{\text{pump}}$. Next, a standard lock-in measurement of the longitudinal resistance $R_{xx}$ is performed at an excitation current $I_{\text{sd}} = 3$ nA, low enough to avoid extra nuclear polariza-tion. Subsequently, with the current switched off, an rf signal is added to the back-gate bias via a bias tee for a duration of $\tau_{\rm rf} = 10$ s for the purpose of generating the NER. $R_{xx}$ is then remeasured to record its difference following the NER. The resistively-detected NER spectra for $^{75}$As nuclei are obtained by repeating the above sequence while varying the frequency of the applied rf signal. For reference, the resistively-detected NMR spectra are also taken using rf irradiation from a triple-turn coil surrounding the sample.

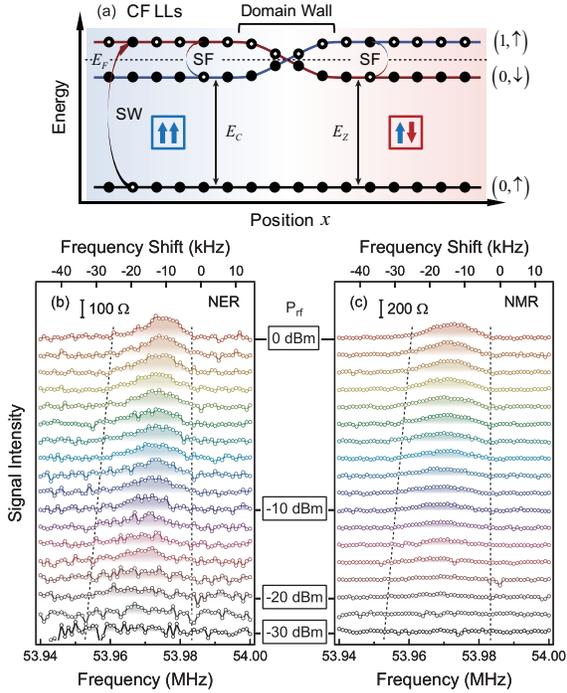

FIG. 1: (a) Schematic energy level diagrams of CF LLs characterized by orbital and spin degrees. The dotted line denotes the Fermi level for the CFs. With respect to the DW, the right-side and left-side regions represent the UD(↑↓) and PD(↑↑), respectively. (b) and (c) Resistively-detected NER and NMR spectra taken with varying rf power $P_{\rm rf}$. Each plot was collected after nuclear polarization was prepared with $\tau_{\rm pump} = 20$ s and $I_{\rm pump} = 100$ nA. The frequency shift from $f_0$ offers the upper axes. Vertical scales differ between (b) and (c). The dotted lines are guides to the eye to the peak narrowing in both spectra.

Figures 1(b) and 1(c) display the spectral comparison of the resistively-detected NER and NMR obtained over a range of different rf power $P_{\rm rf}$. Peaks develop in both spectra from $P_{\rm rf} = -30$ dBm. The peak broadening under smaller $P_{\rm rf}$ is due to different magnitudes of Knight shift occurring in each domain. Under larger $P_{\rm rf}$, the NER and NMR spectra exhibit a peak narrowing. Here, in the case of the NER, the experimental parameter of $P_{\rm rf}$ can be converted into the effective rf-modulated filling factor $\nu_{\rm rf}$ [20]. Figure 2(a) concurrently plots the integrated intensity of the NER and NMR spectra against $P_{\rm rf}$ (upper axis) or $\nu_{\rm rf}$ (middle axis). The NER intensity ($n = 1$ for the fundamental resonance) surpasses the NMR intensity only for $P_{\rm rf} \lesssim -10$ dBm. More noteworthy is that the NER intensity displays saturation behavior with increasing $P_{\rm rf}$ or $\nu_{\rm rf}$ while the NMR intensity continues a monotonic increase despite identical DNP conditions. This is attributed to the fact that the NMR takes place at any position satisfying its resonance condition whereas the NER is restricted to local regions exposed to the DW oscillation. In addition, the NER is produced at subharmonic resonances since the DW oscillation creates in-plane hyperfine fields containing harmonic components. The signal intensity of the subharmonic NER ($n = 2$ and 3) is plotted in Fig. 2(a). In spite of an overall signal suppression, the saturation characteristic seems to be maintained for $n = 2$. Spectral density of the oscillating hyperfine fields declines at higher order of $n$, thereby rendering the NER intensity much weaker.

The spatial fluctuation of nuclear polarization locally modifies $E_Z$, relocating the filling factor required for the SPT away from $\nu = 2/3$. The largely distorted sites serve as pinning centers for the DW, and the DW depinning requires $\nu_{\rm rf}$ to change beyond a critical filling factor of $\nu_c$ [22]. This value is here determined from the net nuclear polarization in the domains. The lower panel of Fig. 2(a) shows the NER intensity as a function of $\nu_{\rm rf}$ scaled by $\nu_c$, where $\nu_{\rm rf}/\nu_c$ below or above unity classifies the DW displacement into strong and weak pinning regimes. The intensity evolution is indicative of growth saturating into a plateau, signaling these two regimes. Moreover, as $\nu_c$ becomes smaller for lower nuclear polarization, the DW oscillation can interact with nuclear spins beyond the pinning centers. In Fig. 2(b), the NER signal evolution is verified by changing $\tau_{\rm pump}$. Although the signal strength also depends on the degree of the nuclear polarization, the saturation behavior is found to remain robust over both regimes. Additionally, a great change in signal size for different $I_{\rm pump}$ rules out the possibility of nuclear polarization being attained in thermal quasi-equilibrium during $\tau_{\rm pump}$. Figure 2(c) presents the temperature dependence of the saturation behavior. The observed signal evolution is considerably quenched around $300 - 400$ mK due to the thermal instability of the domain structures. This supports that the detected signal originates from the interplay between the DWs and the nuclear spins.

In order to simulate the NER response, a one-dimensional model is introduced along the direction of the DW motion. Via the filling-factor modulation, the DW oscillates with a total amplitude of $A_{\rm DW}$ [Fig. 3(a)]. The DW width, $W$, is theoretically estimated to be $W_0 \sim 4l_B$ ($l_B$: magnetic length) in the ground state [21]. $M_x(x)$ and $M_z(x)$ stand for the in-plane and out-of-plane magnetizations in the DW, respectively. The NER response is calculated by integrating over all QW-matrix nuclei feeling the local Knight shifts $K(x, z)$ [22];

$$I(f) \propto \int_{-\infty}^{\infty} dx \int_{-w/2}^{w/2} dz\, g(f; x, z) S_n(x) |P_N(x)|, \quad (1)$$

where $g(f; x, z) = \exp\left\{-[f - f_0 + K(x, z)]^2/\Gamma^2\right\}$ is a single spin packet with the bare frequency $f_0 = 53.986$ MHz and the nuclear dipole broadening $\Gamma = 2$ kHz. The $n$-th order spectral density $S_n(x)$ is obtained from the Fourier transform of the $M_x$-component oscillation. As seen in Fig. 3(b), the lowest odd-order spectral density $S_1(x)$ is then localized in the proximity of $x = \pm A_{\text{DW}}/2$. In contrast, since during $\tau_{\text{pump}}$ the nuclear polarization is considered to diffuse away from the fixed DW position ($x \simeq 0$), its distribution is represented by a Gaussian function of $|P_N(x)| = |P_{N,0}| \exp\left[-(|x| - W/2)^2/4D\tau_{\text{pump}}\right]$ [23], with the nuclear polarization degree $|P_{N,0}|$ and the nuclear diffusion coefficient $D$ in the domains [Fig. 3(c)]. Considering that the product of $S_n(x)$ and $|P_N(x)|$ in Eq. (1) mainly determines the resulting signal, $A_{\text{DW}}$-dependent investigation of the NER highlights the spatial profile of $|P_N(x)|$.

The simulated NER spectrum based on $|P_N(x)|$ at $D/l_B^2 = 10$ s$^{-1}$ is shown in Fig. 3(d), which seemingly reproduces the slightly asymmetric lineshape observed in Fig. 1(b). In Fig. 3(e), the calculated spectra are mapped out as a function of $A_{\text{DW}}$. With increasing $A_{\text{DW}}$, the peaks related to the PD and UD approach each other, displaying a trend similar to the spectral narrowing observed in Fig. 1(b). However, these facts are in common with the NMR spectra seen in Figs. 1(c) and 3(d). Therefore, the NER spectral shape is likely to not reflect the $|P_N(x)|$-shape decaying on the order of the nuclear diffusion length [e.g. Fig. 3(c)], but is most likely to be governed by electron heating in our present system.

The spatially varying Knight shifts are responsible for the differing NER frequencies, which means that the individual nuclear spins react only at the resonant frequency even though exposed to the localized spectral density. In order to treat a full nuclear response, the spectral integration is further taken with respect to the frequency. The experimental results presented in Figs. 2(a)-2(c) can then be compared with the simulation results at different $D/l_B^2$, shown in Fig. 2(e). Here $A_{\text{DW}}$ is assumed to be parametrically characterized by the scaled $\nu_{\text{rf}}/\nu_c$. Evidently, the experimentally observed saturation behavior cannot be explained simply by the calculation based on the dipolar-dominated diffusion coefficient, $D_i/l_B^2 \sim 0.1$ s$^{-1}$ (i.e. $D_i \sim 10$ nm$^2$/s) [1, 17]. The saturation trend is instead in agreement with the calculations using a larger $D$. In particular, this $D$ reproduces the widely expanded plateau in the weak pinning regime [see Fig. 2(b)]. Also, in Fig. 2(d) the intensity evolution with $n$ closely resembles that seen in Fig. 2(a). At elevated temperature, the DW configurations are expected to smooth out due to the thermal spin excitation across the narrow energy gap of the SPT. This implies that the DW width is effectively regarded as $W = \gamma(T) \times W_0$ with a temperature-dependent coefficient $\gamma \geq 1.0$. Figure 2(f) shows the calculation results at several $\gamma$. The signal strength is primarily influenced by $|P_{N,0}|$, which changes with temperature according to relevant experimental values given in Ref. [22]. In addition, the signal rise becomes more moderate as the DW width increases with $\gamma$. These facts are consistent with the thermal tendency of the intensity evolution in Fig. 2(c).

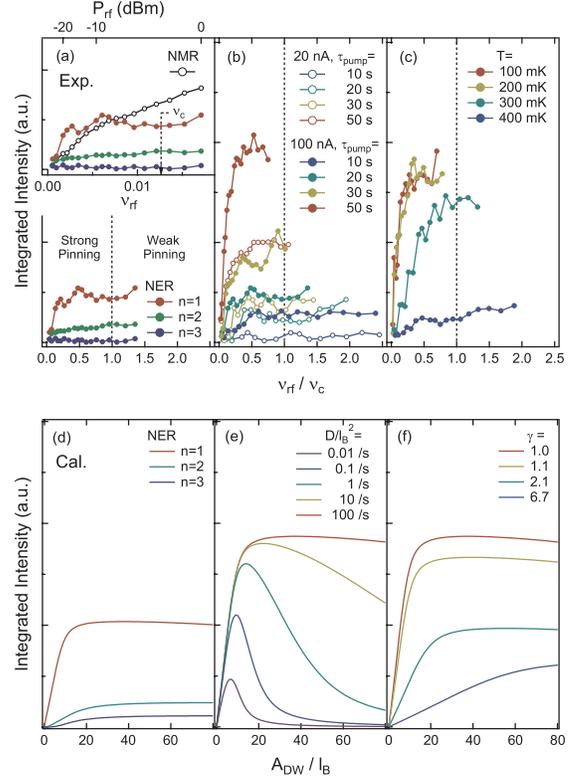

FIG. 2: (a) Upper panel: experimentally obtained evolution of the integrated NER and NMR intensity as a function of the effective rf-modulated filling factor $\nu_{\text{rf}}$ (middle axis) or the rf power $P_{\text{rf}}$ (upper axis) for $\tau_{\text{pump}} = 20$ s and $I_{\text{pump}} = 100$ nA. The NER intensity evolution is also shown for different subharmonic order $n$. Lower panel: the NER integrated intensity replotted for the scaled $\nu_{\text{rf}}/\nu_c$. (b) Pumping-time dependence of the NER intensity evolution for $I_{\text{pump}} = 20$ nA and 100 nA. (c) Temperature dependence of the NER intensity evolution at the fixed $\tau_{\text{pump}} = 50$ s ($\tau_{\text{init}} = 10$ s) and $I_{\text{pump}} = 100$ nA. (d)-(f) Calculated evolution of the integrated NER intensity (d) for various order of $n$, (e) for different magnitude of $D/l_B^2$, and (f) for several temperature-dependent coefficients $\gamma$. $D/l_B^2 = 100$ s$^{-1}$ is used for the simulations in the panels of (d) and (f).

The simulated DW movement is assumed to be free from the pinning centers. Nevertheless, the simulations provide qualitative agreement on the distinct scales of $\nu_{\text{rf}}/\nu_c$ and $A_{DW}$. This comparison assigns a simple estimation of the pinning center range $L_{\text{pin}} \sim 30l_B (\approx 280$ nm). In parallel, it is found that the nuclear polarization is delivered over the maximum investigated distance, $d_{\max} = |\pm A_{DW}/2| \sim 40l_B (\approx 380$ nm), away from the DWs during the short-time $\tau_{\text{pump}} = 10$ s. Since $d_{\max}$ is less than half the reported domain dimension of $3-10\ \mu$m [24], the pumped nuclear polarization quickly approaches a homogeneous distribution around the DWs. The diffusion coefficient for nuclear spins is then estimated to be at least $D_e/l_B^2 \gtrsim 100$ s$^{-1}$ (i.e. $D_e \gtrsim 10^4$ nm$^2$/s). Comparing with the intrinsic value of $D_i$, the nuclear diffusion is apparently enhanced in the presence of the domains.

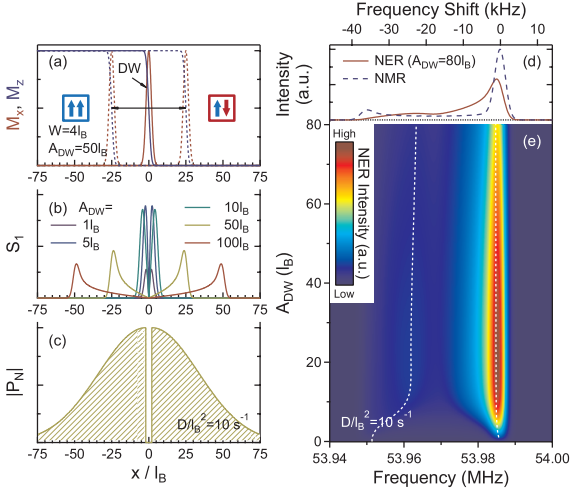

FIG. 3: (a) Modeled in-plane and out-of-plane magnetization components, $M_x$ and $M_z$, across the DW separating the PD ($\uparrow\uparrow$) and UD ($\uparrow\downarrow$). The dotted lines show the two components at odd quarter oscillations ($W = 4l_B$ and $A_{DW} = 50l_B$). (b) Spectral density for the fundamental resonance $S_1$ extended on the $x$ axis when the DW oscillates with different $A_{DW}$. (c) Spatial distribution of the nuclear polarization near the DW expected for the nuclear diffusion coefficient $D/l_B^2 = 10$ s$^{-1}$ ($\tau_{\text{pump}} = 50$ s). (d) NER spectrum simulated at $A_{DW} = 80l_B$. For reference, the NMR spectrum calculated assuming $S_1(x) =$ const. and no DW oscillation is shown integrally normalized. The frequency shift from $f_0$ gives the upper axis. The left and right peaks are derived from the Knight shift in the PD ($\uparrow\uparrow$) and UD ($\uparrow\downarrow$), respectively. (e) Simulated map of the NER spectra as a function of $A_{DW}$. The two peak positions are marked by the dotted lines.

The enhancement of nuclear diffusion in the domains can be accounted for as follows: (i) since the electron spin polarization is highly uniform in the PDs and UDs except for the DWs, most of nuclei in the domains feel the uniform Knight fields, distinct from the nuclei in the quantum dot. As the nuclear Zeeman splitting is comparable among nuclei in each domain, the energy conservation for nuclear spin flip-flop makes angular momentum transfer between nuclei more frequent; (ii) In addition, nuclear spin transport in the domains may occur via quasiparticle-quasihole pairs i.e. spin excitons [25]. The neutral quasiparticle of the spin exciton propagates as a spin-flip (SF) mode or spin-wave (SW) mode. At $\nu = 2/3$, the SF transition between $(1,\uparrow)$ and $(0,\downarrow)$ is allowed both in the UDs and PDs [26], while the SW excitation between $(0,\uparrow)$ and $(0,\downarrow)$ exists mainly in the PDs, as depicted in Fig. 1(a). By inelastic light scattering measurements, the ferromagnetic SW mode has been in fact observed in the spin-polarized phase at $\nu = 2/3$ [27], and the SF mode has been affirmed below the SW mode energy [28].

In principle, such spin excitons can be created via the hyperfine interaction with a nucleus located at a given position $\mathbf{R}$ [25]; $\hat{H}_{ex-N}(\mathbf{R}) \propto \sum_{\mathbf{k}} W(k^2)(\hat{A}_{\mathbf{k}}^+\hat{I}^+ + \hat{A}_{\mathbf{k}}\hat{I}^-) \exp[i(k_x X - k_y Y)]$, where $\hat{A}_{\mathbf{k}}$ is the spin exciton operator and $\hat{I}^\pm$ are the transverse components of the nuclear spin operator. Since the energy requirement for the single flip-flop process is removed by the virtual nature of spin excitons, a combined process makes possible an effective nuclear spin-spin interaction, which is in analogy with the Ruderman-Kittel-Kasuya-Yoshida (RKKY) interaction in metals. The *direct* dipole-dipole coupling generally arises between neighboring nuclei at an atomic-scale distance whereas the above *indirect* mechanism allows nuclear spin exchange over a distance in the order of $l_B$. Such long-range coupling provides a possible explanation for a few orders of magnitude higher than $D_i$. Indeed, the coefficient for the spin-exciton mediated diffusion is estimated to be $D_{sd} \approx d^2/\tau_{sd} \sim 4 \times 10^4$ nm$^2$/s according to Ref. [25], when the effective interaction length and the nuclear flip-flop time in the domains are $d = \frac{l_B}{2}\sqrt{\frac{E_C}{E_Z}} \sim 4.7$ nm ($E_C \approx E_Z$ around the SPT) and $\tau_{sd} \simeq 0.01 B^{-\frac{3}{2}}$ s $\sim 0.5$ ms ($B$ in tesla), respectively [22].

On the other hand, as the electron spin polarization gradually transits at the domain boundary, the Knight fields vary inside the DWs, where the Zeeman splitting mismatch between nuclei reduces the nuclear flip-flop probability. Since on the basis of Figs. 1(b) and 1(c) the global Knight shift is $\Delta_K \sim 30$ kHz between the PDs and UDs, the local Knight shift differs by $\delta_K (\approx a\Delta_K/\sqrt{2}W_0) \sim 0.3$ kHz between nearest-neighboring homo-nuclei in the DWs, where $a = 0.565$ nm is the lattice constant of GaAs. By comparing $\Delta_K$ and $\delta_K$ with $\Gamma$, it is suggested that the *indirect* flip-flop interaction between distant nuclei is inactive across the DWs while the *direct* dipole-dipole coupling between neighboring nuclei is still active in the DWs. Hence the DWs partially act as a barrier to block mutual nuclear diffusion between the UDs and PDs. These arguments give an account for the unintuitive fact that positive and negative nuclear polarization separately exists in the respective domains without canceling out each other [22].

In conclusion, the magnetic resonance tracing based on the NER revealed that the nuclear polarization is rather homogeneously distributed beyond the pinning centers on the periphery of the DWs. The enhanced diffusion has never been reported in the experiments performed at an integer filling $\nu = 2$ [29], where the creation of spin excitons in the second LL is prohibited by a large energy gap of the cyclotron energy. In contrast, in the fractional QH ferromagnet at $\nu = 2/3$, the CF cyclotron energy is compensated by the CF Zeeman energy. Therefore, the spin excitons are more likely to be generated to mediate the nuclear diffusion. This *indirect* exchange coupling is identified in the past study on nuclear spin decoherence [30], and may provide a key way to communicate remote nuclear spin information [31]. Moreover, the RKKY interaction among nuclear spins has been studied as a driving force for nuclear ferromagnetic ordering at a millikelvin range [32, 33]. Hence the understanding of the *indirect* exchange coupling provides a significant tool for making sense of a novel state of strongly correlated nuclear spins.

We would like to thank K. Muraki and NTT Basic Research Laboratories for supplying high quality wafers for this study. We are grateful for valuable discussions with J. N. Moore, G. Yusa, K. Akiba, M. H. Fauzi, P. Stano, and L. S. Vlasenko. The reported experiments were carried out in the ERATO Nuclear Spin Project supported by JST. Y.H. acknowledges support from WPI-AIMR at Tohoku University and MEXT KAKENHI Grant Numbers 15H05867 and 26287059.

Supplemental Material for
"NMR Tracing of Hyperfine-Mediated Nuclear Spin Diffusion
in Fractional Quantum Hall Domain Phases"

S. Miyamoto,[1,2] T. Hatano,[1] S. Watanabe,[3] and Y. Hirayama[1]

[1]*Department of Physics, Tohoku University, 6-3 Aramaki Aza Aoba, Aoba-ku, Sendai 980-8578, Japan*
[2]*School of Fundamental Science and Technology, Keio University,*
*3-14-1 Hiyoshi, Kohoku-ku, Yokohama 223-8522, Japan*
[3]*Institute of Science and Engineering, Kanazawa University, Kanazawa 920-1192, Japan*
(Dated: May 23, 2016)

### A. Additional remarks on NER model calculation

For the NER model calculation, the in-plane and out-of-plane magnetizations of the domain wall (DW) are assumed to be $M_x(x) = \exp\left[-4\ln(2)(x-x_0)^2/W^2\right]$ and $M_z(x) = \left[1 - \mathrm{sgn}(x-x_0)\sqrt{1-M_x^2}\right]/2$, respectively [1]. The DW located at a position of $x_0$ sinusoidally oscillates with a total amplitude of $A_\mathrm{DW}$, where the DW oscillation incorporates no distortion stemming from any pinning centers. The Knight shift is given by $K(x,z) = \alpha_\mathrm{full} P_e(x) |\psi_e(z)|^2$, taking into account a spatial variation of spin/charge density both in the $x$ and $z$ directions. The Knight-shift magnitude expected for full electron spin polarization is represented by $\alpha_\mathrm{full} = 0.37$ MHz·nm for $\nu = 2/3$ at $B = 7.4$ T [2]. $P_e(x)$ represents the time-averaged electron spin polarization that is deformed by the $M_z$-component oscillation [see Fig. S1(a)], and $|\psi_e(z)|^2 = (2/w)\cos^2(\pi z/w)$ is the probability density of the wave function in the quantum well (QW). In addition, the unshifted resonance frequency $f_0$ and the nuclear dipole broadening $\Gamma$ are determined from the NMR reference for $^{75}$As nuclei recorded under a depletion condition, where the Knight shift is taken to be zero. By contrast, since the electron wavefunction in the QW also introduces a spatial variation in the Knight shift, the nuclear diffusion tends to be slowed down in the $z$ direction. Additionally, the diffusion coefficient in the AlGaAs barrier layer has been known to be on the order of less than 1 nm$^2$/s due to the quadrupole disorder [3–5]. For these reasons, the nuclear polarization hardly leaks from the QW in our experimental conditions. In the model calculation, the nuclear polarization distribution, $P_N(x)$, is simply considered for the $x$ direction, and the average degree of nuclear polarization in the domains, $|P_{N,0}|$, is given as its scaling factor, which is experimentally determined following the procedure detailed in Supplemental Material C. Furthermore, the spectral density, $S_n(x)$, generated by the $M_x$-component oscillation, is calculated for different subharmonic orders $n$ as shown in Fig. S1(b). This supports the fact mentioned in the main text that the lower odd-order $S_n(x)$ is the largest spectral density acting on the nuclei near $x = \pm A_\mathrm{DW}/2$.

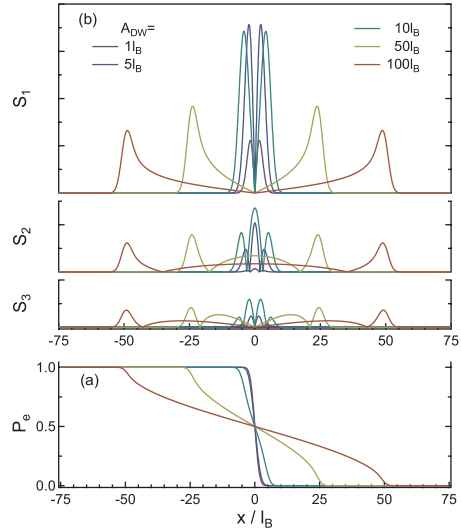

FIG. S1: (a) Time averaged electron spin polarization in the oscillating DW and (b) spatial profile of the calculated $n$-th order spectral density when varying the DW-oscillation amplitude, $A_\mathrm{DW}$, in units of the magnetic length $l_B$.

## B. Finite-thickness effect on spin phase transition and hyperfine-mediated nuclear diffusion

The composite fermion (CF) cyclotron energy is described by $E_C = \frac{2e\hbar}{m_{\text{CF}}} B \left|\nu - \frac{1}{2}\right|$, while the CF Zeeman energy is given as $E_Z = |g_{\text{CF}}|\mu_B(B + B_N)$ including the nuclear magnetic fields $B_N$. The g-factor of CFs is assumed to be the typical value for electrons, $g_{\text{CF}} \sim -0.44$, and the CF effective mass is $m_{\text{CF}} = \xi\sqrt{B}m_e$, which corresponds to the polarization mass for describing the Coulomb energy difference between distinct spin-polarized states [6]. Here, the finite-thickness electron wavefunction in the QW softens the Coulomb potential in the $z$ direction. When one incorporates the finite-thickness effect, the Coulomb potential at in-plane coordinate $r$ can be reasonably approximated as the form of $V(r) \propto (r^2 + \lambda^2)^{-1/2}$, where $\lambda$ represents the effective thickness of the electron wavefunction [7, 8]. The CF cyclotron energy is then reduced to $E_C = \frac{2e\hbar}{\xi m_e}\sqrt{\frac{B_\lambda B}{B+B_\lambda}}\left|\nu - \frac{1}{2}\right|$ with $B_\lambda = \frac{\hbar}{e\lambda^2}$. The CF level coincidence occurs when $E_Z = jE_C$, where $j = 1$ and 2 for $\nu = 2/3$ and 3/5, respectively [9]. The transition points of $(\nu_{\text{tr}}, B_{\text{tr}})$ therefore follow $B_{\text{tr}} = \left[\frac{4j}{\xi'|g_{\text{CF}}|}\left(\nu_{\text{tr}}^{\text{unpol.}} - \frac{1}{2}\right)\right]^2$ in the unpolarized case of $B_N \simeq 0$.

Figure S2(a) represents the $(\nu, B)$-plane map of longitudinal resistance $R_{xx}$ taken in the presence of negligible nuclear polarization. As marked by the open circles, the crossing of the CF levels elicits the SPT-related peaks inside the $R_{xx}$ minima of $\nu = 2/3$ and 3/5. The SPT peaks thus divide the spin-polarized and spin-unpolarized (partially spin-polarized) states on the higher and lower $B$ sides, respectively. For $\nu = 2/3$, the circle plot is fitted with the aforementioned relation centered at $B = 7.4$ T using the effective mass parameter, $\xi' \left(\equiv \xi\sqrt{\frac{B+B_\lambda}{B_\lambda}}\right) \sim 0.51$ T$^{-1/2}$. From the intrinsic value of $\xi \approx 0.4$ T$^{-1/2}$ obtained in Ref. [8], we can estimate $B_\lambda \sim 12$ T or $\lambda \sim 8$ nm. Here, the spin-exciton mediated nuclear diffusion addressed in the main text is considered to be governed by the length scale of $\lambda$ and the effective interaction length $d$ [10]. As the present work corresponds to the case that $\lambda$ is somewhat greater than $d$, the nuclear flip-flop time of $\tau_{sd} \simeq 0.01 B^{-\frac{3}{2}}$ s is employed to calculate the nuclear diffusion coefficient in the domains.

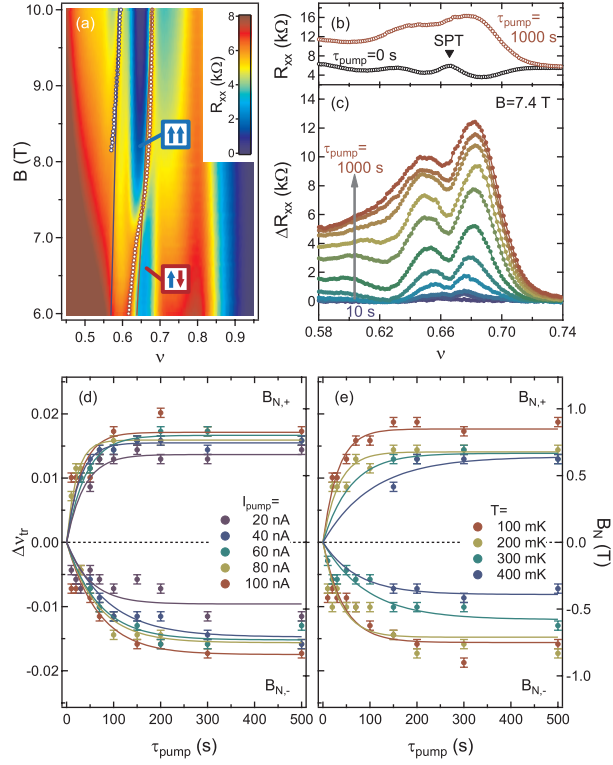

FIG. S2: (a) Color plot of $R_{xx}$ as a function of $\nu$ and $B$. A lower current of $I_{\text{sd}} = 10$ nA is used to suppress the current-induced DNP. The SPT-related peaks of $\nu = 2/3$ and 3/5 are emphasized by the open circles, and the respective solid lines follow the relation described in the supplemental text. The SPT for $\nu = 2/3$ occurs between the UD ($\uparrow\downarrow$) and PD ($\uparrow\uparrow$) phases. (b) SPT peak around $\nu = 2/3$ right after the skyrmion-mediated initialization ($\tau_{\text{pump}} = 0$ s) and the current-induced polarization ($\tau_{\text{pump}} = 1000$ s) at the pumping current of $I_{\text{pump}} = 100$ nA. (c) Evolution of $\Delta R_{xx}$ for the enhanced nuclear polarization with $\tau_{\text{pump}}$. (d) and (e) $\Delta\nu_{\text{tr}}$ (left axis) and $B_N$ (right axis) as a function of $\tau_{\text{pump}}$ for the varied $I_{\text{pump}}$ and different temperatures. The solid lines show the fitting results by an exponential curve.

## C. Bidirectional nuclear spin polarization and domain-wall pinning

Sweeping the filling factor in the vicinity of the SPT leads to the continuous displacement of DWs through the system. The DWs "pick up" local nuclear spin polarization, which enables us to carry out nuclear magnetometry [8, 11]. More specifically, the degree of nuclear spin polarization created in the domains can be extracted from the behavior of the SPT peaks. Since the transition point at a certain $B$ is shifted from $\nu_{\text{tr}}^{unpol.}$ to $\nu_{\text{tr}}^{pol.}$ in the polarized case of $|B_N| > 0$, the nuclear magnetic fields can be accordingly estimated from the filling-factor shift of the transition point;

$$B_N = \frac{4j}{\xi' |g_{\text{CF}}|} \sqrt{B} \cdot \Delta\nu_{\text{tr}}, \tag{S1}$$

where $\Delta\nu_{\text{tr}} \equiv \nu_{\text{tr}}^{pol.} - \nu_{\text{tr}}^{unpol.}$. Here, we focus on the SPT peak for $\nu = 2/3$ ($j = 1$) and fix the magnetic field to $B = 7.4$ T, as used in the main work. Figure S2(b) shows the $R_{xx}$ curves monitored right after the skyrmion-mediated initialization ($\tau_{\text{pump}} = 0$ s) and the current-induced polarization ($\tau_{\text{pump}} = 1000$ s). The dynamic nuclear polarization (DNP) at $\nu_{\text{pump}} \sim 2/3$ leads to a significant change in the global shape of $R_{xx}$ around $\nu = 2/3$ as well as around $\nu = 3/5$. Along the lines of Ref. [12], the curve after $\tau_{\text{pump}}$ is analyzed by subtracting the original curve at $\tau_{\text{pump}} = 0$ s, so that the enhanced part in $R_{xx}$ is separated from the background signal with respect to the SPT peaks. Figure S2(c) shows the DNP-derived signal developed at various $\tau_{\text{pump}}$; $\Delta R_{xx} = R_{xx}(\tau_{\text{pump}}) - R_{xx}(\tau_{\text{pump}} = 0$ s$)$. As the SPT peaks are broadened by the DNP with increasing $\tau_{\text{pump}}$, the positions of the split peaks oppositely branch from $\nu_{\text{tr}}^{unpol.}(= 2/3)$. The peak shifts are measured for different $I_{\text{pump}}$, and plotted as $\Delta\nu_{\text{tr}}$ in Fig. S2(d). Moreover, the peak growing can be ascertained up to 450 mK, at which the SPT peaks are mostly vanishing due to the domain disappearance. Figure S2(e) plots the peak shifts of $\Delta\nu_{\text{tr}}$ extracted for different temperatures below 400 mK. Using Eq. (S1), $\Delta\nu_{\text{tr}}$ yields $B_N$ on the right vertical axes for Figs. S2(d) and S2(e). Thus, the current-induced DNP in the domains produces oppositely directed nuclear magnetic fields denoted by $B_{N,+}(> 0)$ and $B_{N,-}(< 0)$. Such a bidirectional nuclear polarization is in contrast with the case of the optical pumping around the SPT that causes a unidirectional nuclear polarization [13]. The broadened SPT peak, therefore, reflects the fact that positive and negative nuclear polarization is selectively created in the unpolarized domains (UDs) and the polarized domains (PDs), respectively [14]. Then, the average magnitude of nuclear magnetic fields across all domains is $|B_N| = (|B_{N,+}| + |B_{N,-}|)/2$, where the absolute value is used because depolarizing $B_{N,+}$ and $B_{N,-}$ by the NER results in the same signed change in $R_{xx}$ [see Fig. S2(c)]. Note $|B_N|$ reaches as large as 5.3 T in case that all nuclear species are fully polarized in GaAs bulk [15]. Thus, the nuclear polarization degree can be extracted, and input as $|P_{N,0}|$ in the NER model. Meanwhile, the spatial fluctuation of nuclear polarization causes local shifts in the SPT, creating large and small pinning sites for the DWs. The overall shifts of the SPT yield the critical filling factor $\nu_c$ that divides the DW movement into strong and weak pinning regimes. Therefore, in order for the DW oscillation to escape the pinning centers, the effective rf-modulated filling factor $\nu_{\text{rf}}$ needs to exceed $\nu_c = \Delta\nu_{\text{tr},+} - \Delta\nu_{\text{tr},-}$, where the positive and negative sign of $\Delta\nu_{\text{tr},\pm}$ corresponds to $B_{N,\pm}$ in Eq. (S1).